\documentclass[epj-spec,final]{svjour}
\usepackage{amsmath,amssymb,graphicx,bm}

\newcommand{\lm}{\Lambda}
\newcommand{\lb}{\Lambda_{\rm b}}
\newcommand{\be}{\begin{equation}}
\newcommand{\ee}{\end{equation}}
\newcommand{\vnn}{V_{\rm NN}}
\newcommand{\vlowk}{V_{{\rm low}\,k}}
\newcommand{\fmi}{\, \text{fm}^{-1}}
\newcommand{\mev}{\, \text{MeV}}
\newcommand{\kev}{\, \text{keV}}

\begin{document}

\title{Helium halo nuclei from low-momentum interactions}

\author{S.\ Bacca\inst{1}\fnmsep\thanks{\email{bacca@triumf.ca}} 
\and A.\ Schwenk\inst{1}\fnmsep\thanks{\email{schwenk@triumf.ca}} 
\and G.\ Hagen\inst{2}\fnmsep\thanks{\email{hageng@ornl.gov}} 
\and T.\ Papenbrock\inst{2,3}\fnmsep\thanks{\email{tpapenbr@utk.edu}}}

\institute{TRIUMF, 4004 Wesbrook Mall, Vancouver, BC, V6T 2A3, Canada 
\and 
Physics Division, Oak Ridge National Laboratory, P.O.\ Box 2008, Oak Ridge, 
TN 37831, USA 
\and
Department of Physics and Astronomy, University of Tennessee, Knoxville, 
TN 37996, USA}

\abstract{We present ground-state energies of helium halo nuclei
based on chiral low-momentum interactions, using the 
hyperspherical-harmonics 
method for $^6$He and coupled-cluster theory for $^8$He,
with correct asymptotics for the extended halo structure.}

\maketitle

\section{Motivation}

The physics of strong interactions gives rise to new structures in
neutron-rich nuclei. One prominent example are the helium
halo nuclei, $^6$He and $^8$He, with two or four loosely-bound
neutrons forming an extended halo around the $^4$He core. $^6$He
is the lightest halo nucleus and the lightest Borromean system
in nature. Recently, a combination of nuclear and atomic physics 
techniques enabled a new era of precision measurements of the 
ground-state energies (masses)
and charge radii of $^6$He~\cite{Dilling,%
radius6He} and $^8$He~\cite{radius8He,8Heexp}. 
Their reproduction poses
extraordinary challenges for nuclear theory that will advance
{\it ab-initio} methods and our understanding of nuclear forces.

The existing {\it ab-initio} calculations with traditional
nucleon-nucleon (NN) and three-nucleon (3N) potentials are based
on the Green's Function Monte Carlo (GFMC) method~\cite{GFMC} 
and the No-Core Shell Model (NCSM)~\cite{NCSM3N}. In addition,
there are larger-scale NCSM results but restricted to NN 
interactions~\cite{NCSMHe} and Fermionic Molecular Dynamics studies
based on a unitary-correlated NN interaction plus a two-body
potential introduced to mimic 3N effects~\cite{FMD}.

One of the central advances in nuclear theory has been the development
of effective field theory (EFT) and the renormalization group (RG)
to nuclear forces. While
light nuclei have been investigated using the NCSM~\cite{NCSMchiral},
there are no results for helium halo nuclei based on chiral NN and
3N interactions. This is due to the challenges of the loosely-bound
halo and the extended structure of the wave function. In this paper,
we present results of an effort to study helium halo nuclei 
based on chiral EFT. These combine the RG evolution to low-momentum
interactions with the {\it ab-initio} hyperspherical-harmonics method
for $^6$He and coupled-cluster theory for $^8$He.
Our work goes beyond the previous investigation~\cite{Hag06} of the
helium isotopes by studying the cutoff variation, as a tool to probe
the effects of many-body forces, and we present first results based
on the exact hyperspherical-harmonics expansion for $^6$He, which is
more difficult to describe in coupled-cluster theory due to its 
open-shell nature~\cite{Hag06}.

\section{Effective field theory and the renormalization group 
for nuclear forces}
\label{EFTRG}

Nuclear interactions depend on a resolution scale, which we denote
by a generic momentum cutoff $\lm$, and the Hamiltonian is always
given by an effective theory for NN and corresponding many-nucleon
interactions~\cite{pionless,chiral,Vlowk}:
\be
H(\lm) = T + \vnn(\lm) + V_{\rm 3N}(\lm) + V_{\rm 4N}(\lm) + \ldots \,.
\label{Hamiltonian}
\ee
For most nuclei, the typical momenta are of order of the pion mass,
$Q \sim m_\pi$, and therefore pion exchanges are included explicitly
in nuclear forces. In chiral EFT~\cite{pionless,chiral}, nuclear 
interactions are organized in a systematic expansion in powers of
$Q/\lb$, where $\lb$ denotes the breakdown scale, roughly $\lb \sim
m_\rho$. At a given order, this includes contributions from one- or 
multi-pion exchanges and from contact interactions, with short-range
couplings that depend on the resolution scale $\lm$ and for each 
$\lm$ are fit to data. Chiral EFT enables a direct connection
to the underlying theory of Quantum Chromodynamics (QCD) through 
full lattice QCD simulations~\cite{lattice}. This can constrain
long-range pion-nucleon couplings, the pion-mass dependence of nuclear
forces, and has the potential to access experimentally difficult
observables, such as three-neutron properties.

\begin{figure}[t]
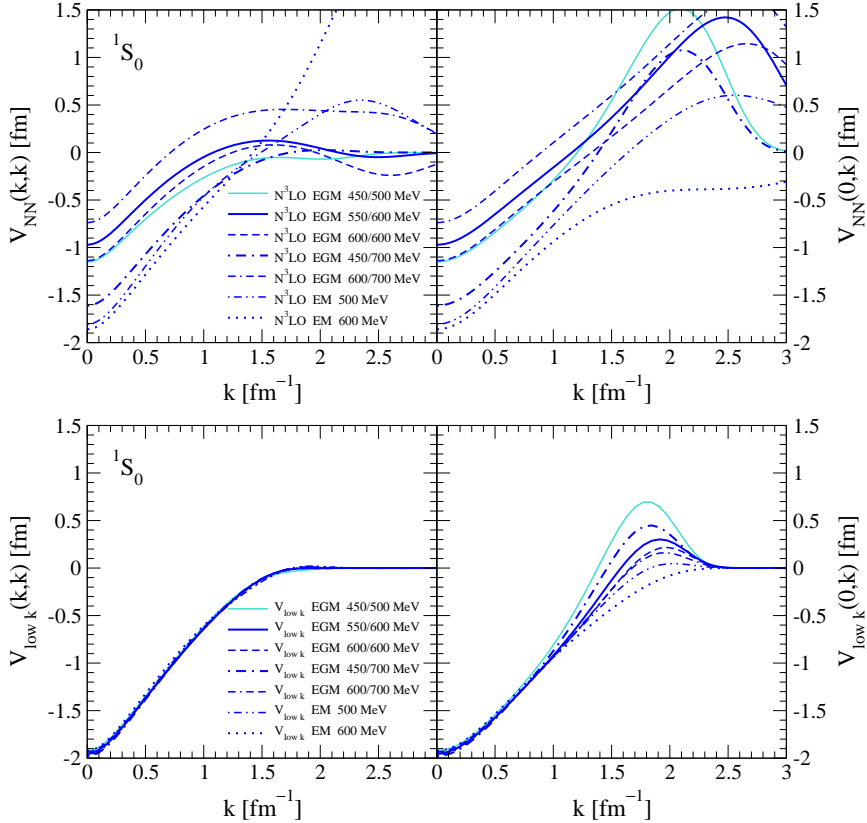

\begin{center}
\includegraphics[scale=0.41,clip=]{vnn_chiral_1s0.eps} \\[1mm]
\includegraphics[scale=0.41,clip=]{vlowk_chiral_1s0.eps}
\end{center}
\caption{Diagonal (left) and off-diagonal (right) momentum-space 
matrix elements of different chiral EFT interactions at 
N$^3$LO~\cite{EM,EGM} in the $^1$S$_0$ channel (upper figures)
and after RG evolution to low-momentum interactions $\vlowk$ 
(lower figures) for a smooth regulator with $\Lambda = 2.0 \fmi$
and $n_{\rm exp} =4$.\label{universality}}
\end{figure}

In Fig.~\ref{universality}, we show chiral EFT interactions at 
N$^3$LO of Entem and Machleidt~\cite{EM} (EM with $\Lambda = 500$
and $600 \mev$) and of Epelbaum {\it et al.}~\cite{EGM} (EGM with 
$\Lambda = 450$--$600 \mev$ and a spectral-function cutoff in 
the irreducible two-pion exchange $\Lambda_{\rm SF} = 500$--$700 
\mev$). These accurately reproduce low-energy NN scattering. Using
the renormalization group (RG)~\cite{Vlowk,VlowkPLB,smooth}, 
we can change
the resolution scale in chiral EFT interactions and evolve N$^3$LO
potentials to low-momentum interactions $\vlowk$ with lower cutoffs.
The RG preserves long-range pion exchanges and
includes subleading contact interactions, so that NN scattering 
observables and deuteron properties are reproduced~\cite{smooth}.
In the lower part of Fig.~\ref{universality}, we show the 
universality of $\vlowk$ by evolving all seven N$^3$LO potentials
to a lower cutoff $\Lambda = 2.0 \fmi$, and that the RG 
evolution weakens the off-diagonal coupling between low and
high momenta. This decoupling can also be achieved using similarity
renormalization group (SRG) transformations towards 
band-diagonal~\cite{SRG1} or block-diagonal~\cite{SRG2} interactions 
in momentum space.

\begin{figure}[t]
\begin{center}
\includegraphics[scale=0.38,clip=]{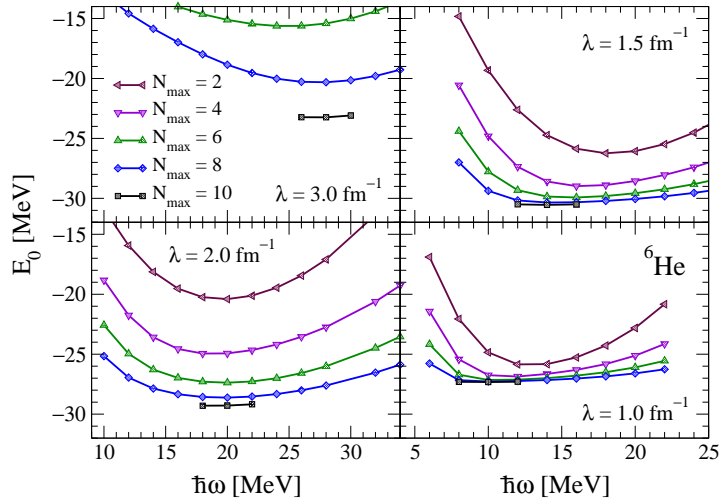}
\end{center}
\caption{Ground-state energy $E_0$ of $^6$He versus oscillator
parameter $\hbar\omega$ for different SRG-evolved interactions with
$\lambda = 3.0, 2.0, 1.5$ and $1.0 
\fmi$. The NCSM results clearly show improved 
convergence with the maximum number of oscillator quanta $N_{\rm max}$ for
lower cutoffs. Since 3N interactions are neglected, the different NN
calculations converge to different ground-state energies.
For details see Ref.~\cite{NCSM}.\label{He6}}
\end{figure}

Changing the cutoff leaves observables unchanged by construction, but
shifts contributions between the interaction strengths and the sums 
over intermediate states in loop integrals. The evolution of chiral 
EFT interactions to lower cutoffs is beneficial, because these
shifts can weaken or largely eliminate sources of nonperturbative
behavior such as strong short-range repulsion and short-range tensor 
forces~\cite{Born}. Lower resolution needs smaller bases in many-body 
calculations, leading to improved convergence in nuclear structure
applications. This is demonstrated by the very promising convergence
for $N_{\rm max} \sim 10$ in NCSM calculations with SRG 
interactions~\cite{NCSM}, shown in Fig.~\ref{He6}.

Chiral EFT interactions become more accurate with higher orders
and the RG cutoff variation provides an estimate of the theoretical
uncertainties due to neglected many-body interactions in $H(\lm)$
and due to an incomplete many-body treatment (see also 
Refs.~\cite{INT,nucmatt}). For 
example, when three-nucleon (3N) interactions are neglected, we 
have found a universal correlation between the $^3$H and $^4$He 
binding energies~\cite{Vlowk3N}, empirically known as ``Tjon-line''.

\section{Challenges for ab-initio calculations of halo nuclei}

The extended structure and the asymptotic behavior of the wave 
function are theoretically challenging for halo
nuclei, and to date there are no results based on chiral NN and 
3N interactions. However, advances in {\it ab-initio} 
methods have the potential to overcome this challenge.

The hyperspherical-harmonics (HH) method has recently been extended
to studies of $^6$He based on simple phenomenological NN 
potentials~\cite{He6Sonia1,He6Sonia2} using a powerful 
antisymmetrization algorithm~\cite{Nir}. The HH wave function is
expanded in a translationally-invariant Jacobi basis and has the 
correct asymptotic behavior. In addition, the method is capable
to handle nonlocal potentials, by expanding the interaction in 
harmonic-oscillator matrix elements~\cite{HHJISP,HHUCOM}.

Coupled-cluster (CC) theory~\cite{Bar07}
is a powerful method for nuclei for which 
a closed-shell reference state provides a good starting 
point~\cite{Kow04,Gour06}. The CC wave function is developed
in a single-particle basis, which can be a Hartree-Fock or
Gamow-Hartree-Fock basis~\cite{Hag06} with correct asymptotic
behavior. Combined with rapid convergence for low-momentum 
interactions, CC theory has pushed the limits of accurate 
calculations to medium-mass nuclei and set new benchmarks 
for $^{16}$O and $^{40}$Ca~\cite{VlowkCC}.

Three-nucleon interactions are a frontier in the physics of
nuclei~\cite{Tokyo} and including their contributions in {\it
ab-initio} calculations of neutron-rich and heavier nuclei
presents a central challenge. For the helium isotopes, 3N 
interactions are crucial for binding energies and radii, 
for the evolution of nuclear structure with isospin, and
for spin-orbit effects (see for example Refs.~\cite{Hag06,%
Vlowk3N,Sofia}). In chiral EFT without explicit Deltas, 3N 
interactions start at N$^2$LO~\cite{chiral3N1,chiral3N2} and 
typically constitute $\sim 10 \%$ of the NN potential 
energy~\cite{INT}. Their contributions are given diagrammatically by
\be
\vspace*{-2mm}
\includegraphics[scale=0.54,clip=]{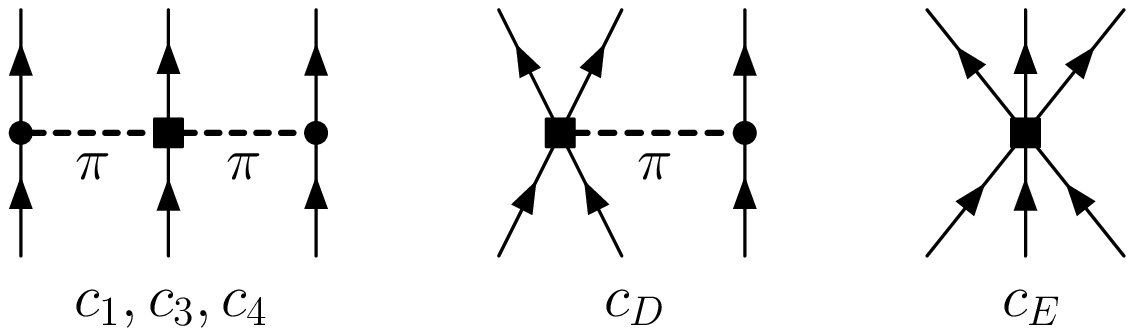} \nonumber
\ee
The long-range two-pion-exchange part is determined by the 
couplings $c_1, c_3, c_4$, which have been constrained in 
the $\pi$N and NN system, and the remaining D-~and E-term
couplings are usually fitted to the $^3$H binding energy 
and another observable in $A \geqslant 3$. The leading
chiral 3N interaction generally improves the agreement
of theory with experiment in light nuclei~\cite{NCSMchiral}.
At the next order, N$^3$LO, there are no new
parameters in chiral EFT for many-body forces.

Since chiral EFT is a complete 
low-momentum basis, we have constructed 3N interactions 
$V_{\rm 3N}(\lm)$, corresponding to RG-evolved interactions, 
by fitting the leading D-~and E-term couplings to the $^3$H binding 
energy and the $^4$He binding energy~\cite{Vlowk3N} or 
radius~\cite{new3N} for a range of cutoffs. By constraining the 3N
interaction with the $^4$He radius, we have found an improved
cutoff dependence of nuclear matter and empirical saturation
within theoretical uncertainties~\cite{new3N}. For lower
cutoffs, low-momentum 3N interactions become 
perturbative in light nuclei~\cite{Vlowk3N}, and the first CC results 
with 3N forces show that low-momentum 3N interactions are 
accurately treated as effective 0-, 1- and 2-body terms, 
and that residual 3N forces can be
neglected~\cite{CC3N}. This is very promising for
developing tractable approximations to handle many-body
interactions in {\it ab-initio} approaches.

\section{Results for helium halo nuclei based on chiral 
low-momentum interactions}

In this Section, we present results for ground-state energies of 
helium nuclei based on chiral low-momentum NN interactions.
These combine the RG evolution to low-momentum interactions
and the resulting improved convergence with the {\it ab-initio}
hyperspherical-harmonics method for $^6$He and coupled-cluster 
theory for $^8$He. Work towards including 3N interactions is 
in progress.

\begin{table}[t]
\begin{center}
\begin{tabular}{ll}
\hline\noalign{\smallskip}
Method & $E_0$($^4$He) [MeV] \\
\noalign{\smallskip}\hline\noalign{\smallskip}
Faddeev-Yakubovsky (FY)~\cite{Andreas} & $-28.65(5)$ \\
Hyperspherical harmonics (HH) & $-28.65(2)$ \\
CCSD (CC with singles and doubles) & $-28.44$ \\
$\Lambda$-CCSD(T) (CC with triples 
corrections~\cite{Bar07,triples}) & $-28.63$ \\
\noalign{\smallskip}\hline
\end{tabular}
\end{center}
\caption{Ground-state energy $E_0$ of $^{4}$He based on the {\it ab-initio}
FY, HH and CC approaches and the two-nucleon $\vlowk$ interaction evolved
from the EM $500 \mev$ chiral N$^3$LO potential~\cite{EM} for a cutoff 
$\Lambda = 2.0 \fmi$ using a smooth $n_{\rm exp} =4$ regulator.
\label{benchmark}}
\end{table}

We first benchmark the HH and CC methods against the exact
Faddeev-Yakubovsky (FY) $^4$He ground-state energy in 
Table~\ref{benchmark}. Our results are based on the low-momentum
NN interaction $\vlowk$
evolved from the EM $500 \mev$ chiral N$^3$LO 
potential~\cite{EM} for a cutoff $\Lambda = 2.0 \fmi$ using
a smooth $n_{\rm exp} =4$ regulator. The 
(variational) HH ground-state energy
agrees very well with the exact FY result, and the CC energy 
also agrees with the FY result after triples corrections are 
included. At the CCSD level (CC theory with singles and doubles
excitations), only about $200 \kev$ of correlation energy is
missed. We have found similar agreement for all other cutoffs
studied in this work.

\begin{figure}[t]
\begin{center}
\includegraphics[scale=0.4,clip=]{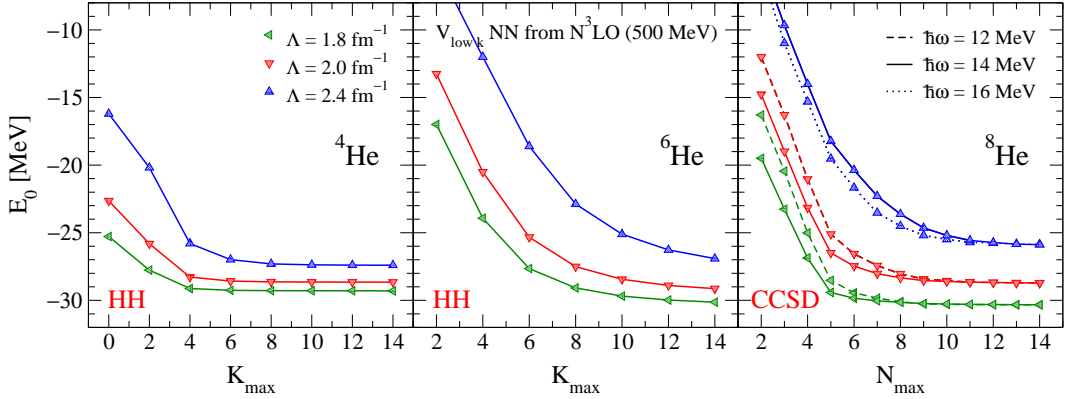}
\end{center}
\caption{Ground-state energies $E_0$ of $^{4}$He, $^{6}$He and $^{8}$He for
low-momentum interactions $\vlowk$ evolved from the EM $500 \mev$ chiral
N$^3$LO potential~\cite{EM} for cutoffs $\Lambda = 1.8$, $2.0$ and $2.4 
\fmi$ using a smooth $n_{\rm exp} =4$ regulator. We show the convergence of
the hyperspherical-harmonics (HH) results for $^{4}$He and $^{6}$He as a
function of the grand-angular momentum $K_{\rm max}$ and the (CCSD level)
coupled-cluster results for $^{8}$He as a function of the number of
oscillator shells $N_{\rm max} = {\rm max}(2n+l)$.\label{4-8He}}
\end{figure}

In Fig.~\ref{4-8He}, we show the convergence as a function of basis
size for the HH $^{4}$He and 
$^{6}$He ground-state energies and the CC $^8$He ground-state energy
based on low-momentum NN interactions $\vlowk$ for a range of cutoffs
$\Lambda = 1.8$, $2.0$ and $2.4 \fmi$. The HH convergence as a 
function of the grand-angular momentum $K_{\rm max}$ is excellent for
the $^{4}$He energies for all cutoffs studied, whereas the $^6$He
energies are not completely converged, but as expected, we find improved 
convergence for lower cutoffs. The HH results practically do not 
show a dependence on the oscillator parameter $\hbar \omega$ introduced
by the expansion of nonlocal potentials in harmonic-oscillator 
matrix elements~\cite{HHUCOM}. For the CC $^8$He energies,
we obtain a good convergence with the number of oscillator shells 
$N_{\rm max} = {\rm max}(2n+l)$ using the spherical CC 
code~\cite{CCspherical}. The CCSD results are obtained in a Hartree-Fock
basis and exhibit a small $\hbar\omega$ dependence,  which decreases
with $N_{\rm max}$.

\begin{table}
\begin{center}
\begin{tabular}{llll}
\hline\noalign{\smallskip}
$\Lambda$ [fm$^{-1}$] &
$E_0$($^4$He) & $E_0^{\infty}$($^6$He) [$E_0(K_{\rm max}=14)$] & 
$E_0$($^8$He) $\Lambda$-CCSD(T) [CCSD] \\
\noalign{\smallskip}\hline\noalign{\smallskip}
$1.8$ & $-29.30(2)$ & $-30.28(3)~~\,[-30.13]$ & $-31.21~[-30.33]$ \\
$2.0$ & $-28.65(2)$ & $-29.35(13)~[-29.13]$ & $-29.84~[-28.72]$ \\
$2.4$ & $-27.40(2)$ & $-27.62(19)~[-26.91]$ & $-27.54~[-25.88]$ \\
\noalign{\smallskip}\hline\hline\noalign{\smallskip}
experiment & $-28.296$ & $-29.268$~\cite{6Heexp} & $-31.395$~\cite{8Heexp} \\
\noalign{\smallskip}\hline
\end{tabular}
\end{center}
\caption{Ground-state energies $E_0$ in MeV of $^{4}$He (HH converged),
$^{6}$He (HH extrapolated and for the largest $K_{\rm max}=14$ space) and
$^{8}$He ($\Lambda$-CCSD(T) and CCSD level) based on Fig.~\ref{4-8He}. For
comparison we also give the experimental ground-state energies.}
\label{4-8Hetab}
\end{table}

Our results for the ground-state energies of $^{4}$He, $^{6}$He and 
$^{8}$He are summarized in Table~\ref{4-8Hetab}. For $^6$He, the 
largest $K_{\rm max}=14$ space in the HH calculations includes three
million basis states. The resulting matrices are dense and larger spaces are
hard to achieve using the HH methods, but we are working towards 
accomplishing this and fully-converged $^6$He results. At present,
we make an extrapolation assuming an exponential Ansatz
$E(K_{\rm max}) = E^{\infty} + \alpha \, e^{-\beta K_{\rm max}}$, with fit
parameters $\alpha, \beta$. The results of this extrapolation are 
listed in Table~\ref{4-8Hetab}. Our procedure to obtain $E^{\infty}$ is
based on an extrapolation of the $K_{\rm max}=8-14$ (last four) points.
The first two $K_{\rm max}=2-4$ points are omitted as they contain too
few basis states to be considered as a reasonable expansion of the wave
function. As an estimate of the error, we take double the difference
between the extrapolation of the $K_{\rm max}=8-14$ (last four) and 
$K_{\rm max}=6-14$ (last five) points. We adopted this procedure, since
it was robust when applied to $^4$He. For $^4$He, besides neglecting 
the first two points as explained, we also omitted the $K_{\rm max}=12$
and $K_{\rm max}=14$ results, to simulate a not fully converged energy.
By extrapolating the $K_{\rm max}=4-10$ (last four) points we obtained
$E^{\infty}$ values that agree with $E(K_{\rm max}=14)$ within the
theoretical uncertainty. For $^8$He, we show CC ground-state energies 
at the CCSD level, which typically accounts for $90\%$ of the correlation
energy~\cite{Bar07} and at the $\Lambda$-CCSD(T) level~\cite{triples}.
The energies in Table~\ref{4-8Hetab} correspond to the $N_{\rm max}=14$
results, where a satisfactory convergence is reached.

The cutoff variation of the energies in Table~\ref{4-8Hetab} is
significantly larger than the theoretical uncertainties due to an 
incomplete many-body treatment (the extrapolation errors or
neglected quadrupole and higher corrections). Therefore, the cutoff variation
is almost entirely due to neglected many-body interactions in 
the Hamiltonian $H(\lm)$~\cite{INT,Vlowk3N}. Our results highlight
that 3N interactions are crucial for ground-state energies, and
it is encouraging that the experimental energies are within the
cutoff variation, so within the effects expected from many-body
forces. Even for cutoffs around $2.0 \fmi$, where the
NN-only results are reasonably close to the experimental 
energies for $^4$He and $^6$He, the $^8$He ground-state energy
is underbound. Although low-momentum 3N interactions
are overall repulsive in nuclear~\cite{nucmatt} and neutron 
matter~\cite{nm}, the same two-pion-exchange $c_i$-terms are attractive
in $^4$He for these cutoffs~\cite{INT,Vlowk3N} and the N$^2$LO chiral
3N interaction provides an attractive contribution to spin-orbit
splittings~\cite{NCSMchiral}.

\section{Outlook}

This is an exciting era, with a coherent effort to understand and
predict nuclear systems based on EFT and RG interactions, where 
3N forces are a frontier, and with major advances in {\it ab-initio}
methods for nuclear structure. On the experimental side, a highlight
is set by the recent precision measurements of the charge radii and
the first direct mass measurements of $^6$He~\cite{Dilling,radius6He}
and $^8$He~\cite{radius8He,8Heexp}. However, due to the extended halo
structure, there are no results for the helium halo nuclei based
on chiral NN and 3N interactions.

We have presented results for the ground-state energies of 
helium halo nuclei based on chiral low-momentum NN interactions.
This combines the RG evolution to low-momentum interactions with
the {\it ab-initio} HH method for $^6$He and CC theory for $^8$He.
These approaches overcome the challenges of the extended halo structure and
have the correct asymptotic behavior of the wave function.
The HH and CC methods have been validated against the exact FY 
result for $^4$He. For $^6$He and $^8$He, our results highlight
the importance of 3N interactions. For all studied cutoffs, the
NN-only results underbind $^8$He (see also Ref.~\cite{Hag06}).
Therefore, the helium isotopes
probe 3N effects beyond the overall repulsion in infinite nuclear and 
neutron matter. Work is in progress towards including chiral 3N
interactions in HH and CC approaches.

\begin{acknowledgement}
It is a pleasure to thank the organizers for a very stimulating ENAM08
conference and N.~Barnea, S.~Bogner, D.~Dean, J.~Dilling, B.~Friman, 
R.~Furnstahl and A.~Nogga for many discussions. This work was supported
in part by the Natural Sciences and Engineering Research Council (NSERC),
the U.S.~Department of Energy under Contract Nos.~DE-AC05-00OR22725 with
UT-Battelle, LLC (Oak Ridge National Laboratory), and DE-FC02-07ER41457
(SciDAC), and under Grant No.~DE-FG02-96ER40963 (University of Tennessee).
TRIUMF receives federal funding via a contribution agreement through 
the National Research Council of Canada. This research used resources 
of the National Center for Computational Sciences at Oak Ridge National
Laboratory.
\end{acknowledgement}

\end{document}